# A Novel Privacy-Preserving Deep Learning Scheme without Using Cryptography Component

Chin-Yu Sun[1,*], Allen C.-H. W[1], TingTing Hwan[1]

[1]Department of Computer Science, National Tsing Hua University,

101, Section 2 Kuang Fu Road, Hsinchu, Taiwan, R.O.C.

*Corresponding author
Chin-Yu Sun
sun.chin.yu@gmail.com
Department of Computer Science,
National Tsing-Hua University
TEL:+886-3-571-5131 ext.33573

# A Novel Privacy-Preserving Deep Learning Scheme without Using Cryptography Component

**Abstract**

Recently, deep learning, which uses Deep Neural Networks (DNN), plays an important role in many fields. A secure neural network model with a secure training/inference scheme is indispensable to many applications. To accomplish such a task usually needs one of the entities (the customer or the service provider) to provide private information (customer's data or the model) to the other. Without a secure scheme and the mutual trust between the service providers and their customers, it will be an impossible mission. In this paper, we propose a novel privacy-preserving deep learning model and a secure training/inference scheme to protect the input, the output, and the model in the application of the neural network. We utilize the innate properties of a deep neural network to design a secure mechanism without using any complicated cryptography component. The security analysis shows our proposed scheme is secure and the experimental results also demonstrate that our method is very efficient and suitable for real applications.

## 1 Introduction

Today, deep learning technologies have been widely applied to many segments of applications. These application designers are usually focused on "how to train a useful model" but largely ignore the security issue. This security concern has greatly affected the practicality of the applications. We use the following examples to illustrate the importance of the privacy issue that plays an important role in Artificial Intelligence (AI) applications.

Recently, many EDA companies have applied various AI techniques to their CAD tools for improving the IC-design productivity [1]. For instance, the automatic placement and routing (APR) is an important physical design technology in the back-end flow of the electronic design automation (EDA) design. In the APR, it first performs the floorplanning and placement followed by the routing procedure. The routing procedure first performs a global routing to estimate the overall routing area based on the placement result, and then performs a detailed routing procedure to complete the routing. The correlations between the global route and the detailed route are critical when conducting the "routability analysis" (number of wires in a grid) in the floorplanning and placement stages. If they are not well correlated, it may cause excessive iterations of the global/detailed routing and/or placement/routing procedures that will result in unbearable run times. Hence, some EDA companies intend to use machine learning techniques for better correlation predictions so that their IC-design customers can use this information to achieve an efficient design process. In order to accomplish this mission, the EDA companies need to build up a model (by training) between the global and detailed routing, which is depended upon a database of extensive floorplan/placement information owned by the IC-design customers. However, to the IC-design customers, the floorplan/placement data contains confidential information regarding to their IC products. On the other hand, the EDA companies also will not reveal their models for security reason. Building an effective model requires massive human-labeled data, powerful computing hardware, meticulous parameters setting, and researchers' efforts. Without the mutual trust between EDA companies and IC-design customers, it will be an impossible mission for the current available methods.

The similar dilemmas also occur in some medical applications. Diabetic retinopathy is one of the important causes of blindness. Early diabetic retinopathy usually has no special symptoms, causing people with diabetes often only pay attention to the control of blood sugar while ignoring other physical conditions. If the symptom is detected early and provided with proper treatment, most patients can avoid surgery and maintain vision in a better condition. Recently, ophthalmologists try to use machine learning for better detection of diabetic retinopathy. They use a huge amount of retinopathy images for training a medical model that can precisely label the problematic area from a patient's retina image. Unfortunately, the diabetic information is usually owned by the endocrinology department, not the ophthalmology department. Although the ophthalmologist may have the technology to detect diabetic retinopathy, the medical record (patient's retina image and the analysis result) does not allow to exchange between both departments under some privacy law's protection, such as Patients' Bill of Rights and Health Insurance Portability and Accountability Act.

To resolve the above dilemmas has motivated us to investigate the security issue and develop a privacy-preserving deep learning model. In this paper, we use the innate properties of a deep neural network to design a secure mechanism without using the fully homomorphic encryption, the garbled circuit, or the watermark. Our proposed scheme can protect the privacy of the input, the output, and the model in the application of the neural network, and support the transformation from a non-secure model into a secure model.

The rest of the paper is organized as follows. In Section 2, we present the related work. In Section 3, we give the problem definition. In Section 4, we review the deep learning neural networks. Our proposed scheme is presented in Section 5. After that, an extension idea is presented in Section 6. Our security analysis and experimental results are given in Section 7 and Section 8, respectively. Finally, we conclude the paper in Section 9.

## 2  Related work

In the past, many protecting strategies have been proposed to solve the cooperation's dilemma in the data inference phase, which can be divided into three categories: the homomorphic encryption-based [10-15], the garbled circuit-based [17], and the watermark-based [19, 20].

The homomorphic encryption (or fully homomorphic encryption [9]) allows the computational operation on an encrypted state, and the decrypted result still matches the same computational operation on the plaintext. If the scheme that supports "arbitrary" computational operation on the ciphertext, it is called the fully homomorphic encryption. In 2006, Barni [10] proposed an interactive protocol to protect the privacy for the data owner using homomorphic encryption. In the first layer of the neural networks, the data owner encrypts her/his input data and sends it to the model provider. Then, the server executes by multiplying the encrypted data with the weights of the first layer and sends back to the data owner. After that, the data owner decrypts, applies the activation function, and encrypts the result again and then sends it to the server for the computational operation of the next layer. The above procedures will be repeated layer by layer till the last layer of the neural networks. However, in 2007 and 2008, Orlandi et al. [11] and Piva et al. [12] both pointed out that the weights of neural networks in [10] can be obtained by the data owner. Dowlin et al. [13] proposed the CryptoNets in 2016. The scheme is based on homomorphic encryption but no need for data owner to interact with the server in each layer of the neural networks. In 2017, Chabanne et al. [14] proposed an improved scheme that extended the neural layer of CryptoNet while keeping accuracy at a high level. Further, Liu et al.'s [15] used the homomorphic encryption and the secret sharing technology to design MiniONN that does not need to change the structure of neural network during the training phase. The fully homomorphic encryption based schemes can protect both input' and output's privacies as well as the model privacy because it is built-in on the provider side. However, the homomorphic encryption (or fully HE) is a complexity system [21] and it has some limitations on a practical application [22].

The garbled circuit [16] is a cryptographic protocol that allows two parties jointly compute a function $f(x, y)$ with their secret inputs $x$ and $y$, where the function $f$ should be represented as a Boolean circuit with 2-input gates, such as AND, OR, XOR gates. Both the garbled circuit and the homomorphic encryption are trying to compute on an encrypted (or garbled) state; hence, the core idea is identical but using different technologies. In 2017, Rouhani et al. [17] designed DeepSecure by using the garbled circuit. In their scheme, the operational structure of the neural network has to convert into a Boolean circuit first. Then, the data owner makes the circuit garbled using her or his garbled keys and sends the garbled tables and one of the garbled keys (that corresponding to the data owner's input bit) to the model provider. After receiving the tables and a key, the model provider and the data owner engage in a 1-out-of-2 oblivious transfer (OT) protocol [18] to get another garbled key (that corresponding to the model provider's input bit) from the data owner. Then, the model provider can evaluate the received tables, compute the corresponding encrypted data inference, and send the result back to the data owner. Finally, the data owner can obtain the result by the rest of the garbled keys. Similar to the fully homomorphic encryption schemes, the garbled circuit based schemes can protect both input' and output's privacies. However, the garbled circuit is a Boolean circuit, which means that the complexity of the bitwise operation depends on the structure of the neural network. Moreover, it needs to handle a huge mount encryption keys and symmetric encryption/decryption operations, since handling one-bit operation requires 6 keys, 8 encryptions, 8 decryptions, and support by an extra one OT protocol.

Unlike the homomorphic encryption and the garbled circuit, the watermark technology does not focus on protecting the privacy of input/output data but the model itself. A trusted third party (such as the Court) can extract the watermark from the model, and then confirms the ownership by a verification mechanism. In 2017, Uchida et al. [19] first proposed a framework to embed watermarks in the DNN. Subsequently, Zhang et al. [20] proposed an improved scheme that does not require a model provider to access all the parameters of a stolen model when extracting the watermarks in [19]. The watermark schemes do not protect the input and output's privacies, and only protect model passively i.e., only after the model has been stolen.

## 3  Problem definition

The privacy issues will occur in the inference phase as well as the training phase. For example, in the inference phase, the customers need to protect their input data and output results without revealing their data to outside world. In addition, we need a dataset to train a model, but the data may come from different customers under their protection. Unless the customers abdicate their privacy right, it is hard to collect enough data for training without any privacy protection mechanism. Hence, we can formally define the target DL environment and the three security requirements as:

Our environment has two entities, i.e., a customer (also called data owner in this paper) and a service provider (also called the model provider in this paper). Here, we assume that the service provider already has a trained model and the customer will use this model to inference their data. Therefore, the privacy requirements in this environment can be defined as below.

***Input's privacy***: The service needs the customers to send their data to the service provider for data inference. The input data may be personal, confidential or sensitive in nature. If in the inference phase, it requires to protect the confidentiality of the customers' input data. We say that it is an input privacy protection.

***Output's privacy***: After the data inference, the model will give a predicted result of the corresponding input data. If the results are required to keep confidential, we say that it is an output privacy protection.

***Model's privacy***: A completed model consists of the neural network's topology (structure) and the parameters (values of the weight and the bias) on every neuron. If the model is required to keep confidential, we say that it is a model privacy protection.

## 4  Deep neural networks

Neurons are the basic elements in a deep neural network. Each neuron is connected to a number of neurons around it, thereby to construct the complex "layer" structure. Normally, when the neurons in the same layer are fully connected to the neurons in the next layer, we called it "fully connected (FC) layer". The input nodes provide information from the outside world to the network and are together referred to as the "Input Layer". The collection of hidden nodes forms a "Hidden Layer", which perform computations and transfer information from the input nodes to the output nodes. Then, the output nodes are collectively referred to as the "Output Layer" and are responsible for computations and transferring information from the network to the outside world.

In order to handle the signals from the previous layer, a neuron normally performs two kinds of computation: 1) the linear and 2) the nonlinear computations. According to different input wires, a neuron has assigned different weights and biases on each of it. These weights and biases are constantly updated in the training phase but stabilized when the training is completed. Then, the weights, the biases, and the input signals are calculated together by a linear computation such as Equation (1).

$$y = w \cdot x + b \tag{1}$$

, where $x$ is the input vector, $w$ is the weight matrix, $b$ is the bias vector, and $y$ is the output vector.

After the linear computation, the neuron performs the nonlinear transformation (also called activation function) to model nonlinear behaviors between the input data and the output result. There are various activation functions which are categorized and summarized in [15]. Here, we only introduce the ReLU function and the Sigmoid function which are two common activation functions using in the deep neural networks. As shown in Equation (2), the ReLU function can easily eliminate the value that is less than 0 and the Sigmoid function can keep the output value between the range of 0 to 1.

$$\text{ReLU: } f(y) = \begin{cases} 0, for \ y < 0 \\ y, for \ y \geq 0 \end{cases}$$
$$\text{Sigmoid: } f(y) = \frac{1}{1+e^{-y}} \tag{2}$$

After finishing both the linear and the nonlinear computations, the handled signal is sent to the next neural layer as a new input. Then, it repeats the above procedures layer by layer till the last (output) layer. Finally, we can obtain a predicted result $y'$. If $y'$ is obtained in the inference phase, this result is the final predicted result; otherwise, we then start to perform the backpropagation algorithm [26] by using $y'$ and the correct answer $y$. The backpropagation algorithm is an optimization algorithm for the cost (or called the error) in DNN. The concept of backpropagation is to find the cost at the last layer, which between the correct answer and the predicted result. Then, the algorithm propagates the cost from the back to the front layer to compute the cost for every neuron. Finally, we can minimize the costs by the gradient descent algorithm and optimizes the parameters in DNN. Figure 1 illustrates the architecture of the deep neural networks and an individual neuron.

**Remarks.** We state the proposed scheme is the DNN-based but not convolution neural network (CNN)-based. The reason is that we believe the DNN is the fundamental of the "layer" structure in DL. Due to our proposed scheme can be performed in the "layer" structure, it can be applied to the CNN model easily.

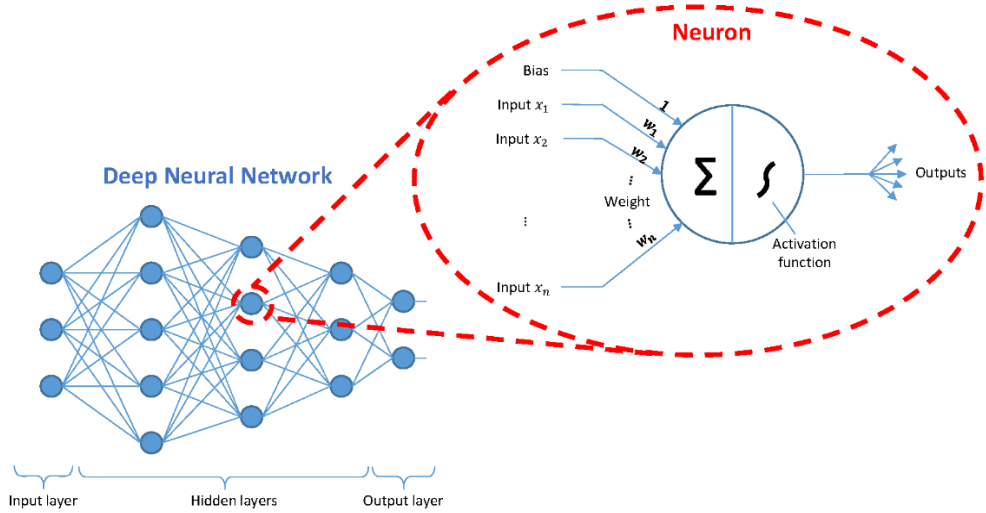

Fig. 1. The DNN architecture and a neuron.

## 5 The proposed scheme

In this section, we propose a novel privacy-preserving scheme which allows two entities (the data owner and the model provider) to securely cooperate by using our secure DNN model. Our scheme is divided into two steps: 1) the complement and 2) the partitioning. The first step fuzzes the output by a complement signal to achieve the effect of encryption. More precisely, we target on the true-false classification (2-classification) learning and try to teach the DNN model to flip its output by a supervised learning method. In the second step, the model provider partitions the model into two independent parts, and then share one part of the model with the data owner. Consequently, the data owner will not obtain the complete model and hence the model provider can protect the model. Furthermore, the data owners can run the model on their sides without revealing the data and thus protect their precious data. We will discuss our proposed scheme in details as follows.

## 5.1 The complement

**(Output's protection)** Because of the model provider owns the whole parameters in the model and usually responsible for the core procedure of the model performing, she or he can easily access to the output data directly. To protect the output's privacy, we defined the first concept of the proposed scheme as follows:

$$d_{out} \oplus cs = F_{[1,M]}(d_{in}||cs) \quad (3)$$

, where $F_{[1,M]}(\cdot)$ is our secure model (with $M$ layer), $d_{in}$ is the input data, $||$ denotes the concatenate operation, $cs \in_R \{0,1\}$ is a complement signal, $d_{out} \in \{True, false\} = \{1, 0\}$ is the predicted (output) result, and $\oplus$ denotes the exclusive OR operation. By observing the Equation (3), the model complements or not its output based on the complement signal $cs$ chosen by the data owner. Since the model provider does not know this $cs$, we say that the output is secure by fuzzing.

## 5.2 The partitioning

**(Input's and model's protection)** Normally, the inference procedure needs interaction between the data owner and the model provider. That is, one of them is relatively disadvantage because she or he needs to provide "something" (whether data or model) to continue the interaction with the other one. To protect both the input data and the model at the same time, we defined the second concept of the proposed scheme as follows:

$$F^c_{[1,k]}(\cdot)||F^s_{[k+1,M]}(\cdot) \xleftarrow{k} F_{[1,M]}(\cdot)$$
$$d^c_{out} = F^c_{[1,k]}(d_{in}||cs) \quad (4)$$
$$d_{out} \oplus cs = F^s_{[k+1,M]}(d^c_{out})$$

, where $\leftarrow$ denotes the partition operation, $k$ is the partitioning layer of the model $F_{[1,M]}(\cdot)$, $F^c_{[1,k]}(\cdot)$ is a client-side model that includes the layers from 1 to $k$ of the secure model $F_{[1,M]}(\cdot)$, $F^s_{[k+1,M]}(\cdot)$ is a server-side model that includes the rest of layers (from $k+1$ to the layer $M$), and $d^c_{out}$ is the output result from the client-side model. The partitioning in Equation (4) means, by given a secure model $F_{[1,M]}(\cdot)$ and a number $k$, the model provider can divide $F_{[1,M]}(\cdot)$ into two subset models: a client-side model $F^c_{[1,k]}(\cdot)$ and a server-side model $F^s_{[k+1,M]}(\cdot)$.

After that, the model provider publishes $F^c_{[1,k]}(\cdot)$ but keeps $F^s_{[k+1,M]}(\cdot)$ secret. Hence, the data owner can input the private data into $F^c_{[1,k]}(\cdot)$ to obtain $d^c_{out}$. Once the data owner sends $d^c_{out}$ to the model provider, the model provider can continue the inferencing procedure. Figure 2 shows the details of the partitioning phase, where the model provider first partitions her/his model $F_{[1,5]}(\cdot)$ into $F^c_{[1,2]}(\cdot)$ and $F^s_{[3,5]}(\cdot)$. Then, the input data $d_{in}$ (ex. a cat image as an input), the complement signal $cs \in_R \{0,1\}$ (ex. 1 as an input; 1 also denotes flip the output), and the client model $F^c_{[1,2]}(\cdot)$ (e.g., the first two layers) are computed by the data owner. After receiving $d^c_{out}$ (the output of $F^c_{[1,2]}(\cdot)$) from the data owner, the model provider then sends $d^c_{out}$ to $F^s_{[3,5]}(\cdot)$ for the rest of the computations. Then, the model provider can obtain a predicted result $d_{out} \oplus cs$ (e.g., dog as an output answer). Finally, the model provider sends $d_{out} \oplus cs$ back to the data owner. When the data owner receives the predicted result from the model provider, she or he can complement the predicted result to obtain the real answer $d_{out}$ (cat as the real answer) according to the chosen signal $cs$.

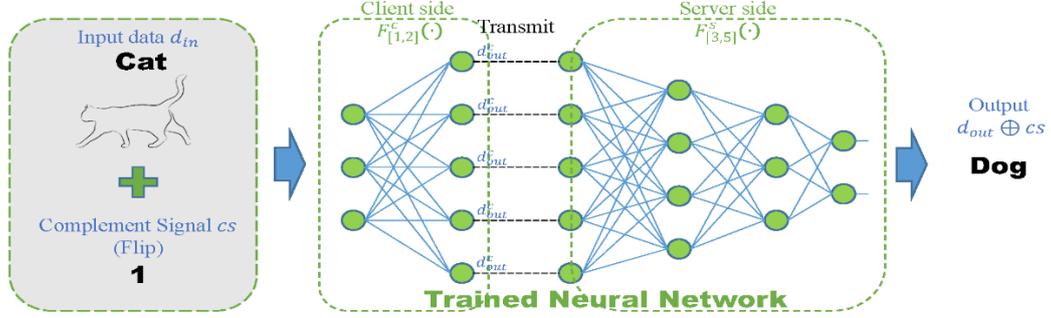

Fig. 2. The partitioning phase of the proposed scheme.

**Remarks.** The data owner only obtains one part of the model. Meanwhile, the model provider cannot extract $d_{in}$ and $cs$ from $d_{out}^c$ (the proof will be shown in Section 7). Therefore, we say that both the input data and the model are secure by the partitioned model.

## 5.3 The secure model

After explaining the two concepts of the proposed scheme, we then introduce how to train a secure model $F_{[1,M]}(\cdot)$ in this subsection. We first defined the concept of the training as follows:

$$L' = NewLabel(D)$$
$$F_s(\cdot) = Retrain(F_{old}(\cdot), k, L', D)$$
(5)

, where $NewLabel(\cdot)$ is a new labeling algorithm, $D$ is a dataset, $L'$ is a new label set, $Retrain(\cdot)$ is our training method, $F_{old}(\cdot)$ is an old (non-secure) model, and $k$ is the partitioned layer of the network in Section 5.2. By observing the Equation (5), the model provider has to perform a new labeling algorithm $NewLabel(\cdot)$ to get the new label set $L'$. Then, she or he uses the dataset $D$ with $L'$ and $k$ to train a secure model $F_{[1,M]}(\cdot)$. Our training method accepts the model provider to reuse their non-secure model $F_{old}(\cdot)$; hence, the $Retrain(\cdot)$ allows the model provider to input $F_{old}(\cdot)$ or nothing.

Now, the model provider can obtain the $L'$ by the new labeling algorithm $NewLabel(\cdot)$. Let the dataset be $D := \{d_i | 1 \leq i \leq m\}$. Then, each element $d_i$ of the dataset $D$ has labeled with the label $l_i \in \{true, false\}$. Now, the model provider randomly generates the complement signal set $CS := \{cs_i | 1 \leq i \leq m\}$, where $cs_i \in \{0, 1\}$. After that, the model provider inputs $D$, $F$, and the labels $l_i$ into the algorithm $NewLabel(\cdot)$. Finally, $NewLabel(\cdot)$ outputs the new label set $L' := \{l_i' | 1 \leq i \leq m\}$, where $l_i' \in \{true, false\}$ is the new label on the element $d_i$. The detail of our label algorithm is shown in Figure 3.

**Algorithm 1** The labeling algorithm $NewLabel(\cdot)$
**Input:** $D$, $CS$, $l_i$
**Output:** $L' : \{l_i'\}$
    **for** each $d_i$ in $D$ **do**
        **if** $cs_i == 0$ && $l_i ==$ true **then**
            sets $l_i' =$ true
        **else if** $cs_i == 0$ && $l_i ==$ false **then**
            sets $l_i' =$ false
        **else if** $cs_i == 1$ && $l_i ==$ true **then**
            sets $l_i' =$ false
        **else**
            sets $l_i' =$ true
        **end if**
    **end for**

Fig. 3. The labeling algorithm

After finishing the labeling, the model provider uses the dataset $D$ with the new label set $L'$ and $k$ to train her/his secure DNN model by the following steps:

***Step 1.*** If the model provider already has a non-secure (old) model $F_{old}(\cdot)$ and $k \neq 1$,

she or he directly jumps to Step 2; otherwise, the model provider (does not have any model) directly jumps to Step 6.

***Step 2.*** The model provider fixes all of the parameters (weights and biases) from layer 1 to layer $k-1$ of the model $F_{old}(\cdot)$.

***Step 3.*** Then, the model provider concatenates the complement signal $cs_i$ with the output of the layer $k-1$ and the new concatenated input keeps the rest of computing from the layer $k$ to the last layer $M$.

***Step 4.*** After that, the model provider uses the output $y'$ with corresponding new label $l'$ then performs the backpropagation algorithm (Section 4).

***Step 5.*** Repeat Step 3 until the training is finished.

***Step 6.*** The model provider concatenates the complement signal $cs_i$ with the input data $d_i$ and computes from the layer 1 to the last layer.

***Step 7.*** Then, the model provider uses the output $y'$ with corresponding new label $l'$ then performs the backpropagation algorithm.

***Step 8.*** Repeat Step 6 until the training is finished.

After that, the model provider can obtain the secure model $F_{[1,M]}(\cdot)$. Figure 4 shows that the difference between a normal (non-secure) DNN model and our proposed (secure) model.

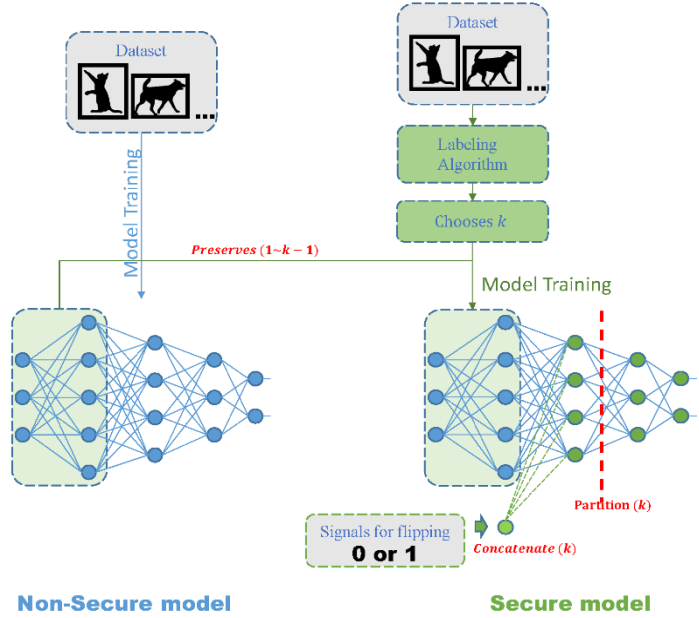

Fig. 4. The difference between the non-secure model and our secure model.

**Remarks 1.** The reason why we separate the model on $k$ is to ensure two input data (such as the $d_{in}$ and $cs$) are already well mixed in layer $k$. Otherwise, the model provider can easily separate two input data by the received result cause they are only concatenating operation (ex. $d_{in}||cs$) before layer $k$.

**Remarks 2.** The number $k$ also decides the tradeoff between the accuracy and the training time of our secure model. When $k$ becomes bigger, the model provider can gain better accuracy and lower retraining time. But in meanwhile, it will leak more parameters of the model to the data owner. The model provider can determine the balance between security level and the performance by our experimental results in Section 8.

# 6  Extension

In this section, we want to introduce an extension idea for upgrading our scheme.

Although our secure scheme now only supports the true/false classification learning (the model's output only true or false, 2-classification), we believe that the main idea in our proposed scheme can easily extend to multiple classification case (the model's output $n$-classification, where $n > 2$). Here, we only briefly explain our extension idea. Similar to the idea that is to teach the model to complement its output by a secure signal, the extension idea becomes to teach the model to shift its output by a secure offset.

We first define an index number on each different kind of output labels in a dataset $D := \{d_i | 1 \leq i \leq m\}$ and each element $d_i$ of the dataset $D$ has labeled with the label $L := \{l_i | 1 \leq i \leq m\}$. Here, the range of $l_i$ is $1 \leq l_i \leq n$, where $n$ as the total number of different kinds of output labels in $D$. Now, the model provider randomly generates the offset $O := \{o_i | 1 \leq i \leq m\}$, where $1 \leq o_i \leq n$. Then, we inputs $D$, $O$, $l_i$, and $T$ into the algorithm $NewLabel\_2(\cdot)$. Finally, $NewLabel\_2(\cdot)$ outputs the new label set $L' := \{c_i' | 1 \leq i \leq m\}$, where $1 \leq c_i' \leq n$ is the new label on the element $d_i$. The detail of the labeling algorithm $NewLabel\_2(\cdot)$ is shown in Figure 5.

---
**Algorithm 2** The labeling algorithm $NewLabel\_2(\cdot)$
---
**Input:** $D$, $O$, $l_i$, and $T$
**Output:** $L' : \{l_i'\}$
    **for** each $d_i$ in $D$ **do**
        computes $l_i' = o_i + l_i \pmod{n}$
    **end for**
---

Fig. 5. Our new labeling algorithm for $n$-classification.

After that, we train the model using eight steps in Subsection 5.3 and separate the model using the partitioning method in Subsection 5.2. Finally, the model should be able to recognize the offset and shift its output by it. Hence, our extension idea should have the ability to protect the input data, the output result, and the model in the multiple-classification learning model.

## 7 Security analysis

This section first presents the Cramer's rule and our definition. Then, we substantiate our system is secure (input's privacy) under the Cramer's rule and give a suggestion for security consideration. After substantiating the privacy of the input, we then prove our secure model can protect both the output's and the model's privacies in the last two subsections.

### 7.1 Cramer's rule and our definitions

Cramer's rule, a theorem in linear algebra, is a formula for the solution of a system of linear equations with as many equations as unknowns.

**Theorem 1** *(Cramer's rule). A matrix of the form* $Ax = b$ *has a unique solution* $x$, ***if and only if*** *the matrix of coefficients* $A$ *is non-singular, where* $A$ *is a* $n \times n$ *matrix,* $x = (x_1, x_2, \dots, x_n)^T$ *and* $b = (b_1, b_2, \dots, b_n)^T$. *The unique solution* $x$ *is given by* $x_i = det(A \leftarrow_i b)/det(A)$, *where* $A \leftarrow_i b$ *is the matrix obtained from* $A$ *by replacing the* $i$-*th column of* $A$ *by the column of constants* $b$.

According to Theorem 1, we can define the unsolvability in Definition 1.

**Definition 1 (Unsolvability)** Let $\mathcal{A}$ be a probabilistic polynomial time adversary, $\mathcal{A}$ cannot obtain the unique solution by solving the linear equations when the number of unknowns is greater than the number of equations.

*Proof.* First, the linear equations with $n$ unknowns and $m$ equations, which described as follows:

$$\begin{aligned} a_{11}x_1 + a_{12}x_2 + \cdots + a_{1n}x_n &= b_1 \\ a_{21}x_1 + a_{22}x_2 + \cdots + a_{2n}x_n &= b_2 \\ &\vdots \\ a_{m1}x_1 + a_{m2}x_2 + \cdots + a_{mn}x_n &= b_m \end{aligned} \qquad (6)$$

is equivalent to a matrix equation of the form $A'x' = b'$, where

$$A' = \begin{pmatrix} a_{11} & a_{12} & \cdots & a_{1n} \\ a_{21} & a_{22} & \cdots & a_{2n} \\ \vdots & \vdots & \ddots & \vdots \\ a_{m1} & a_{m2} & \cdots & a_{mn} \end{pmatrix}, \; x' = \begin{pmatrix} x_1 \\ x_2 \\ \vdots \\ x_n \end{pmatrix}, \text{ and } b' = \begin{pmatrix} b_1 \\ b_2 \\ \vdots \\ b_m \end{pmatrix}. \qquad (7)$$

Obviously, $A'$ is a non-square matrix when $n > m$; hence the determinant of $A'$ never exists. According to Theorem 1, if the matrix of coefficients $A'$ is not a non-singular, then the adversary $\mathcal{A}$ cannot obtain the unique solution. ∎

## 7.2 Input's privacy

In the deep neural networks, the weights and biases are in a fixed size (kernel size), are all calculated together with the input data with the same size of the kernel, by a linear transform that defined in Equation (1). Thus, the computation result of this area will compress into a point (or a value) of the next layer.

According to the number of the kernels in a layer, the input data in the same area may repeatedly compute by different kernels to generate many different output values. Consequently, a set of linear equations can easily be constructed for the same input data as follows:

$$w_{11}x_1 + w_{12}x_2 + \cdots + w_{1n}x_n + b_1 \cdot 1 = y_1 \qquad (8)$$

$$w_{21}x_1 + w_{22}x_2 + \cdots + w_{2n}x_n + b_2 \cdot 1 = y_2$$

$$\vdots$$

$$w_{m1}x_1 + w_{m2}x_2 + \cdots + w_{mn}x_n + b_m \cdot 1 = y_m$$

, where $x$ is an input data (with $n$ size), and $w$ and $b$ are the weights and the biases in $m$ sets of the kernel, respectively.

Now, assume that a model provider $\mathcal{A}$, who also plays the role of an adversary, tries to recover the input data $x = \{x_1, x_2, \ldots, x_n\}$ by receiving results $y = \{y_1, y_2, \ldots, y_m\}$ and all of the kernel (the model provider has all of the kernel). According to Definition 1, $\mathcal{A}$ cannot obtain the unique $x$ when the number of the equation sets $m$ is less than the number of the unknowns $n$. Hence, the separating phase of the proposed scheme is secure, if $n > m$.

However, when we target at a fixed input data, we need to consider the leakage issue caused by shifting the kernel: the number of equations may increase because the DNN algorithm makes the kernel shifts with an amount step until it covers the whole input data. For this reason, when a kernel is overlapped into any pixel of the target image, each stride for the kernel will give the adversary $\mathcal{A}$ more equations (the number of leakage equations is equal to total kernels number per shift). Figure 4 shows that the relationship between a target data and an overlapping kernel on the equation leakage issue.

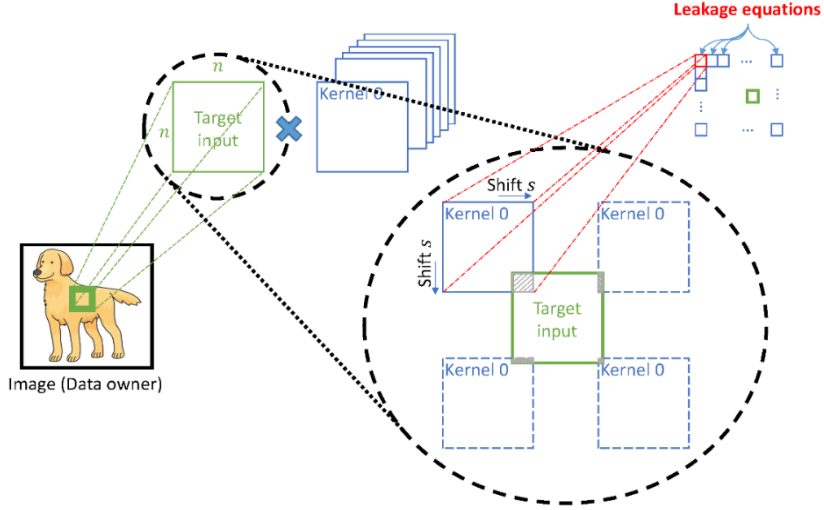

Fig. 4. Example for the equation leakage.

Although the overlap moving may leak more equations, it also generates more unknowns because it covers more input data. Then, we summarize the maximum number of the leakage equations $M$ and the total unknowns $N$, in the worst case, by the following equations:

$$M = (\lceil \frac{2n}{s} - 1 \rceil)^2 \times k$$

$$N = ((n - s) + (\lceil \frac{2n}{s} - 1 \rceil \times s)) \times l \quad (9)$$

, where $n$ denotes the length (or the height) of the kernel, $l$ denotes the width of the kernel, $k$ denotes the number of the kernel, and $s$ denotes the stride size.

## 7.3 Suggestions

Therefore, when separating a model, the model provider needs to use Equation (9) to check $M$ and $N$ carefully such that $N > N$. Take the first layer of AlexNet for example, $N$ equals to 2187 and $M$ equals to 2400. In this case, the adversary $\mathcal{A}$ can eliminate 2400 equations to 2187 equations, in order to obtain the $2187 \times 2187$ square matrix. Therefore, $\mathcal{A}$ has a **non-negligible** of probability to get the unique solution by Theorem 1.

Fortunately is, the equation leakage issue can be solved easily by adjusting the kernel number. As shown in Equations (10), when the model provider changes the number of the original kernel 96 to 64, then $N$ and $M$ become 2187 and 1600, respectively. Hence, due to $N > M$, it becomes secure again.

$$
\begin{aligned}
&N > M \\
&\Rightarrow ((n-s) + (\left\lceil \frac{2n}{s} - 1 \right\rceil \times s)) \times l > (\left\lceil \frac{2n}{s} - 1 \right\rceil)^2 \times k \\
&\Rightarrow \frac{((n-s) + (\left\lceil \frac{2n}{s} - 1 \right\rceil \times s)) \times l}{(\left\lceil \frac{2n}{s} - 1 \right\rceil)^2} > k
\end{aligned} \quad (10)
$$

## 7.4  Output's privacy

Once the input privacy is protected, the privacy of the output can be easily achieved. The reason is that our well-trained model can complement its output by a complement signal from the data owner. As long as the model provider cannot obtain the complement signal from the received result, she or he cannot know the predicted result is complemented or not. Then, the model provider obtains the correct output answer only has a 50% chance. Obviously, it equals to the randomly guessing. Hence, our proposed scheme can protect the privacy of the output.

## 7.5  Model's privacy

The partitioning phase of the proposed scheme can protect the privacy of the model. In this phase, the model provider only publishes a little part of the parameters in the model to the data owner. It is impossible that the data owner reconstructs the whole model without the rest of the parameters. Moreover, the data owner also does not know the topology (neural network's structure) of the model, which also increases the difficulty of reconstruction. Hence, our proposed scheme can protect the privacy of the entire model.

## 8 Experimental results

### 8.1 Experimental setting

In this section, we evaluate the performance and accuracy of our proposed scheme. The experiments were conducted on a machine with an Intel i5-8400 CPU, 32GM RAM, the Ubuntu 16.04 LTS operating system, and a Geforce GTX 1070Ti GPU with 8GB GDDR5. We separately implemented two schemes in Python 3.5 with Tensorflow: 1) The D&C model and 2) the secure model. The D&C model can identify a dog or a cat from an input image. We use this model for simulating a non-secure model. After that, we then apply our proposed scheme to transform the non-secure model to our secure model. Simultaneously, we also test the different preserving layer $N$ for the model accuracy and the corresponding training time. The AlexNet is the basic DNN model in our experimental, and Dogs vs. Cats [2] and MNIST [3] are two datasets used to train it. The configuration of the AlexNet, Dogs vs. Cats, and MNIST are shown in Table 1. For a fair comparison, the configurations of all of the models in this section are under the same setting.

**Table1.** The configurations on DNN and datasets of the proposed scheme.

| DNN and datasets | Configurations |
|---|---|
| AlexNet_v2 | ✓ 5 convolution layers + 3 fully connect layers <br> · Conv1: kernel size : $11 \times 11 \times 3 \times 64$, stide: 4 <br> · Conv2: kernel size : $5 \times 5 \times 64 \times 192$, stide: 1 <br> · Conv3: kernel size : $3 \times 3 \times 192 \times 384$, stide: 1 <br> · Conv4: kernel size : $3 \times 3 \times 384 \times 384$, stide: 1 <br> · Conv5: kernel size : $3 \times 3 \times 384 \times 256$, stide: 1 <br> · FC6: size: 1024 |

|  | · FC7: size: 1024 |
|  | · FC8: size: 2 classes |
|  | ✓ Batch size = 16 |
|  | ✓ Dropout =0.5 |
|  | ✓ Learning rate = 0.001 |
|  | ✓ Power = 0.9 |
|  | ✓ Momentum = 0.9 |
| Dogs vs. Cats | ✓ 10000 dog images + 10000 cat images for training |
|  | ✓ 2500 dog images + 2500 cat images for validation |
| MNIST | ✓ 12665 zero/one images for training and validation |

In the first D&C model, we use the Dogs vs. Cats to train the AlexNet, which is used to pretend the old non-secure model that is to classify the images of dog and cat only. The second model is our secure model, which extends the complement signal to the D&C model. Here, we adopt 0 and 1's images from the MNIST to replace the complement signals 0 and 1 in our secure model. The training method is that we first randomly choose one image (with the size $244 \times 244 \times 3$) from Dogs vs. Cats and the other image (with the size $28 \times 28 \times 1$) from MNIST. Then, we input both images into the model according to our proposed scheme. Here, the dog/cat image should work (together with weights and biases) as usual and the 0/1 image should wait until the concatenating layer. When two images are ready to concatenate with each other, we first resize the 0/1 image according to the size of the output from the dog/cat image and directly concatenate both images as a new input for the next layer. Therefore, the new input that contains original data and the complement signal can easily enter the next layer again then continues its training.

## 8.2 Training time and accuracy

In order to simulate the training time of our secure model, we have to train the D&C model as a base model. Then, we preserve the different layer number of the D&C model to test the training time of our secure model. Table 2 shows the result of the training time. As we retain more layers to practice our secure model, we observe that the completed time begins to decline in the same training steps. In contrast, if we train the secure model without using any layer of the non-secure model (it also means that we directly train the secure model), the training needs more time to finish its job.

Table 2. Comparisons of Training Time (in seconds)

| Training steps | The D&C model | The secure model ($N = 0$) | The secure model ($N = 1$) | The secure model ($N = 2$) | The secure model ($N = 3$) | The secure model ($N = 4$) |
|---|---|---|---|---|---|---|
| 5,000 | 172 | 264 | 143 | 136 | 136 | 126 |
| 10,000 | 323 | 391 | 284 | 270 | 270 | 250 |
| 20,000 | 650 | 694 | 568 | 564 | 541 | 499 |
| 40,000 | 1261 | 1304 | 1130 | 1127 | 1073 | 998 |
| 60,000 | 1965 | 1919 | 1699 | 1618 | 1614 | 1497 |
| 80,000 | 2531 | 2519 | 2264 | 2150 | 2146 | 1997 |
| 100,000 | 3173 | 3137 | 2830 | 2688 | 2690 | 2493 |
| 200,000 | 6280 | 6203 | 5649 | 5379 | 5355 | 4970 |

Second, we tested the accuracy of our secure model. Similarly, we preserve the different layer number of the D&C model in our training and compare each result with the D&C model. To calculate the accuracy rate, we randomly pick 100 images from the validation set and check the number of the correct/wrong answer. Then, we repeat 1000 times and calculate the average accuracy. Our average accuracy (AA) is defined as follows:

$$AA = \frac{\sum_{i=0}^{1000} \left(\frac{c_i}{c_i + ic_i}\right)}{1000} \quad (11)$$

, where $c$ represents the number of the correct predicted answers and $ic$ represents the number of the wrong (incorrect) predicted answers. Table 3 shows the experimental result of the average accuracy. Here, we can observe that it does not work well at the case of $N$ equal to zero if we do not have enough training steps. However, when we start to preserve the layers of the D&C model for training our model, the average accuracy becomes very close to the D&C model.

Table 3. Comparisons of Average Accuracy

| Training steps | The D&C model | The secure model ($N = 0$) | The secure model ($N = 1$) | The secure model ($N = 2$) | The secure model ($N = 3$) | The secure model ($N = 4$) |
|---|---|---|---|---|---|---|
| 5,000 | 0.88054 | 0.67094 | 0.81058 | 0.83932 | 0.8734 | 0.87805 |
| 10,000 | 0.94051 | 0.80123 | 0.88359 | 0.91631 | 0.93036 | 0.93062 |
| 20,000 | 0.9771 | 0.86277 | 0.92085 | 0.93351 | 0.94859 | 0.95807 |
| 40,000 | 0.9816 | 0.88192 | 0.92815 | 0.94402 | 0.95832 | 0.96605 |
| 60,000 | 0.98213 | 0.89313 | 0.93074 | 0.94641 | 0.96176 | 0.96807 |
| 80,000 | 0.98059 | 0.89345 | 0.93863 | 0.9458 | 0.95789 | 0.97242 |
| 100,000 | 0.98356 | 0.89908 | 0.93516 | 0.94244 | 0.96082 | 0.97119 |
| 200,000 | 0.98224 | 0.90333 | 0.93343 | 0.9452 | 0.95766 | 0.97353 |

For an in-depth exploration of the training time, we focus on the convergence rate and compare our secure model with the D&C model. As shown in Figure 5, the convergence rates of our secure model, in different training steps, is still better than the D&C model. Moreover, we also evaluate the convergence rate of our secure model in four different preserving cases as shown in Figure 6, such as preserving the kernel values of (a) the first layer $(N = 1)$, (b) the first two layers $(N = 2)$, (c) the first three layers $(N = 3)$, and (d) the first four layers $(N = 4)$. Even though the performance in Fig. 6(a) is not as good as the original one, the convergence rate becomes fast when we start to preserve the first two, three, and four layers. The results have shown that our conversion method not only works (assists the original model in transmuting) but also fast with good accuracy.

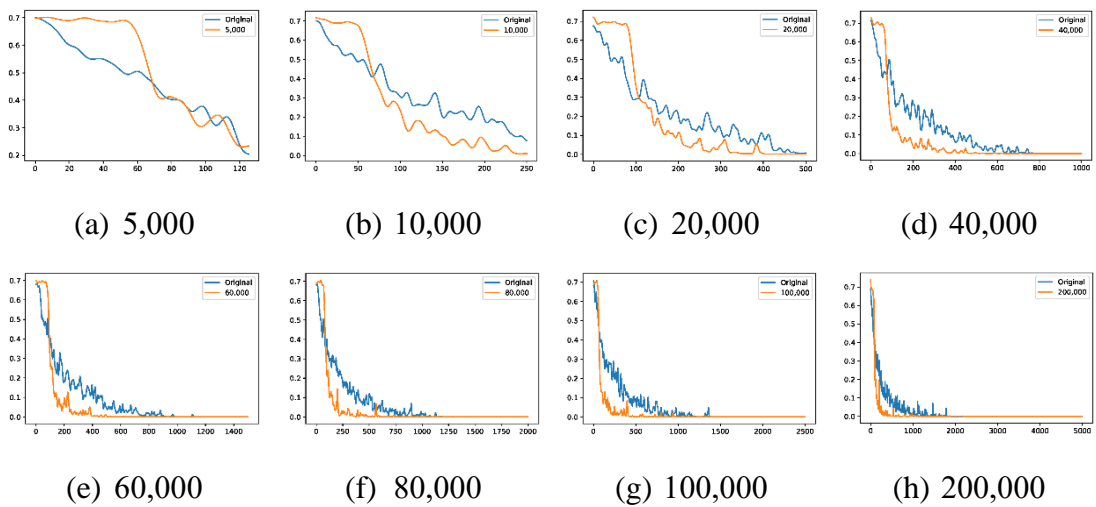

(a) 5,000   (b) 10,000   (c) 20,000   (d) 40,000

(e) 60,000   (f) 80,000   (g) 100,000   (h) 200,000

Figure 5. Convergence time of different training steps of the original mode (D&C model) and our secure model ($N = 3$).

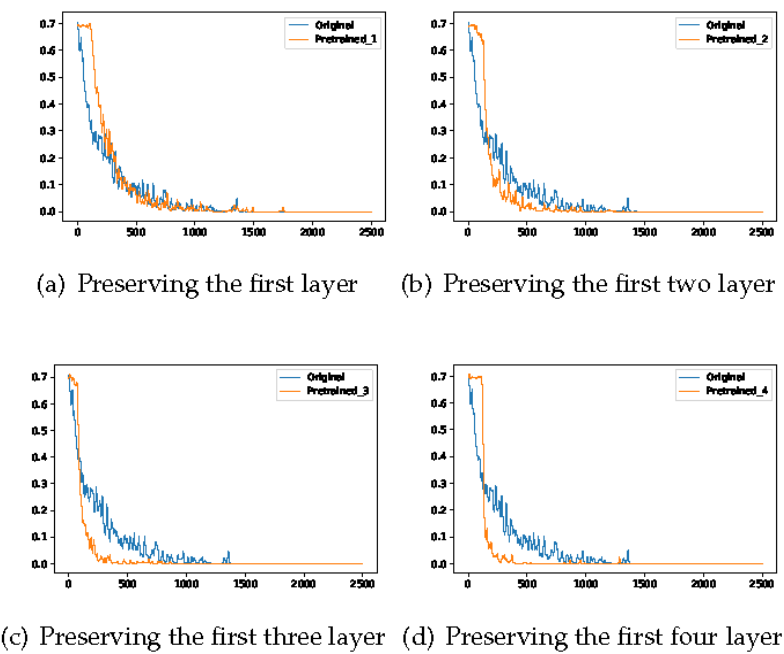

(a) Preserving the first layer   (b) Preserving the first two layer

(c) Preserving the first three layer   (d) Preserving the first four layer

Figure 6. Convergence times for different preserving layers of the secure model.

## 8.3 Storage and communication costs

In order to evaluate the overhead of a data owner who performs our secure scheme, we first simulate the additional storage space that a data owner needed in different partitioning layer of the secure model. Due to the parameters (weights and biases) are all presented in floating point, we adopt the standard of IEEE 754 that double-precision floating-point format. More precisely, we use 64 bits for presenting one floating point number. Hence, we can easily summarize the total parameters (kernel size) needed in different partitioning layer of the secure model and multiplied by 64 bits per parameter to obtain the total storage cost. Table 4 shows the storage space requirement.

**Table 4.** The requirement of the storage space.

| The secure model | Number of parameters | Storage space requirements |
|---|---|---|
| $N = 1$ | 23,296 | $\approx 182$ kB |
| $N = 2$ | 21,164,032 | $\approx 2.52$ mB |
| $N = 3$ | 63,655,936 | $\approx 7.59$ mB |
| $N = 4$ | 148,615,168 | $\approx 17.72$ mB |

kB: kilobyte; mB: megabyte

Then, we evaluate the communication time in an environment of 1Mb per second upload/download speed. In this case, we estimate one link in the output is presenting in 8 bits (according to the feature map image, 8 bits per pixel). Hence, we can use the total links multiplied by 8 bits and then divided by the transmission bandwidth to evaluate the communication times for each partitioning layer of the secure model. The estimated communication times are shown in Table 5.

**Table 5.** The estimated communication times

| The secure model | Number of links | Communication time requirements (millisecond) |
|---|---|---|
| $N = 1$ | 193,600 | $\approx 184.631$ ms |
| $N = 2$ | 186,624 | $\approx 177.979$ ms |
| $N = 3$ | 519,168 | $\approx 61.89$ ms |
| $N = 4$ | 519,168 | $\approx 61.89$ ms |

# 9   Conclusions

In this paper, we presented a privacy-preserving deep learning model and a secure training/inference scheme without using any cryptography-component, which can protect the privacy of the input, the output and the model in the application of the neural network, and also support the transformation from a non-secure model into a secure model. The security analysis has shown that our proposed scheme is secure. Our experimental results have also shown that our proposed scheme is very effective in training time, accuracy, and storage requirements. Hence, the proposed scheme is secure and suitable for the real applications.

This research opens a new frontier for us to further explore the proposed scheme.

Future work includes further improving the inference accuracy, applying to various neural networks, developing various partitioning methods to reduce the communication cost, and developing an effective training mechanism using the proposed scheme.